\documentclass[%
      aps,
      prd,
      floatfix,
      preprintnumbers,
      preprint,
      showpacs,
      nofootinbib,
      tightenlines,
      amsmath
]{revtex4-1}

\usepackage{bm}
\usepackage[dvips]{color,graphicx}
\usepackage{dcolumn}
\usepackage{hyperref}
\hypersetup{colorlinks,%
            citecolor=black,%
            filecolor=black,%
            linkcolor=black,%
            urlcolor=black,%
            pdftex}
\usepackage{latexsym}
\usepackage{float}

\newcommand{\ie}  {{\em i.e.\/} }

\newcommand{\eg}  {{\em e.g.}}
\newcommand{\ud}     {\mathrm{d}}
\newcommand{\gev}    {\:\mathrm{GeV}}

\newcommand{\gevsq}  {\:\mathrm{GeV}^2}
\renewcommand{\Im}{\mathop{\mathrm{Im}}}

\newcommand{\ceps}{\varepsilon}

\newcommand{\eq}[1]{Eq.(\ref{#1})}

\newcommand{\figx}{0.5\textwidth}
\newcommand{\figxx}{0.75\textwidth}

\allowdisplaybreaks[4]
\begin{document}
\title{%
Study of $W^\pm$ and $Z$ Boson Production in Proton-Lead Collisions at the LHC with KP Nuclear Parton Distributions
}

\author{Peng Ru}
\affiliation{School of Physics \& Optoelectronic Technology, Dalian University of Technology, Dalian 116024, China}
\email[]{pengru@mail.dlut.edu.cn}
\author{S. A. Kulagin}
\affiliation{Institute for Nuclear Research of the Russian Academy of Sciences, Moscow 117312, Russia}
\email[]{kulagin@ms2.inr.ac.ru}
\author{R. Petti}
\affiliation{Department of Physics and Astronomy, University of South Carolina, Columbia SC 29208, USA}
\email[]{roberto.petti@cern.ch}
\author{Ben-Wei Zhang}
\affiliation{Key Laboratory of Quark \& Lepton Physics (MOE) and Institute of Particle Physics,
 Central China Normal University, Wuhan 430079, China}
\email[]{bwzhang@mail.ccnu.edu.cn}



\begin{abstract}
\noindent
We present a detailed study of the (pseudo)rapidity distributions of
massive vector bosons produced in $p+p$ and $p+{\rm Pb}$ collisions at the LHC
within the next-to-leading order approximation in perturbative quantum chromodynamics. 
In particular, we discuss the
impact of different cold nuclear matter effects on this process
using the nuclear parton distributions calculated from the microscopic
model developed by Kulagin and Petti (KP). This model was successfully
applied to study nuclear effects in the deep-inelastic scattering
and the Drell-Yan reactions off various (fixed) target nuclei.
Results are compared with the recent CMS and ATLAS $p+{\rm Pb}$ data with
$\sqrt{s}=5.02$\,TeV per two colliding nucleons.
We found an excellent agreement between the predictions of 
the KP model and the recent LHC data on $W^\pm$ and $Z^0$
production in $p+{\rm Pb}$ collisions, including the differential cross sections,
the forward-backward asymmetries, and $W$ charge asymmetry.
We also discuss the sensitivity of the current and future LHC data to the underlying
mechanisms responsible for the nuclear modifications of parton density functions.
\end{abstract}


\maketitle


\section{Introduction}
\label{sec:intro}

The production of massive vector gauge boson in relativistic hadron-hadron
collisions has been extensively studied at $pp$ and $p\bar p$ collisions at the
LHC and the Tevatron and is well understood by the Standard Model (SM) in terms of
perturbative quantum chromodynamics (pQCD)~\cite{Field:pQCD,Ellis:QCD}.
For this reason, $W^\pm$ and $Z^0$ production is commonly considered as a fundamental
candle for SM physics at the LHC, considering the relatively large yields due to the high
center-of-mass energy and luminosity available, as well as the clean experimental signatures.
The $W/Z$ data from $pp(p\bar p)$ collisions at the Tevatron and the LHC also provide
valuable information about the parton density functions (PDFs) of the nucleon in
global QCD fits~\cite{Alekhin:2015cza, Alekhin:2014sya, Harland-Lang:2014zoa, Ball:2014uwa, Lai:2010vv, Gao:2013xoa}.

The recent precision data from the CMS~\cite{Khachatryan:2015hha,Khachatryan:2015pzs}
and ATLAS~\cite{ATLASW,Aad:2015gta} experiments offer the possibility to
extend the study of $W^\pm$ and $Z^0$ boson production to proton-lead collisions
with $\sqrt{s}=5.02$\,TeV per two colliding nucleons at
the LHC~\cite{Vogt:2000hp, ConesadelValle:2009vp, Paukkunen:2010qg, Guzey:2012jp, Kang:2012am, Ru:2014yma, Ru:2015pfa, Albacete:2013ei, Albacete:2016veq}.
Since the QCD factorization theorem~\cite{Collins:1989gx} is expected to hold for nuclei,
we can still describe this process in terms of pQCD, with the corresponding nuclear PDFs for the lead nucleus.
To this end, we recall that PDFs are universal characteristics of the target at high momentum transfer $Q^2$,
which are driven by nonperturbative strong interactions in the considered target.
The leptonic decays of $W/Z$ bosons produced through the Drell-Yan mechanism (DY) are 
of particular interest in this context, 
since they are not modified by the hot and dense medium created in the heavy-ion collisions and
the decay leptons pass through this medium without being affected by the strong interaction.
Furthermore, the intrinsic asymmetry in the $p+{\rm Pb}$ collision system allows one to probe different
Pb fragmentation regions and nuclear parton kinematics by selecting different rapidity values,
\eg, with observables like the forward-backward asymmetries.
The above considerations make the $W/Z$ boson production in $p+{\rm Pb}$ collisions a very good
tool to study nuclear modifications of PDFs and to test the validity of the QCD factorization
for nuclei. It is worth noting that the LHC data provide a unique opportunity
to access the high $Q^2\sim (100\gev)^2$ phase space region, never explored
before by fixed target deep inelastic scattering (DIS), nor by other experiments.

Several phenomenological parametrizations of nuclear parton distributions (NPDF) are available 
in the literature \cite{Hirai:2007sx, Eskola:2009uj, deFlorian:2011fp, Kovarik:2015cma, Khanpour:2016pph}.
Such analyses assume separate nuclear corrections for each parton distribution, 
which are conventionally extracted from global fits to nuclear data including primarily DIS and DY production. 
With the recent availability of data from heavy-ion collisions at RHIC and the LHC, 
additional data sets are included in NPDF 
analyses~\cite{Eskola:2009uj, deFlorian:2011fp, Kovarik:2015cma,Armesto:2015lrg,Kusina:2016fxy}. 
Although these QCD-based studies are useful in constraining nuclear effects for different partons,
they provide limited information about the underlying physics mechanisms responsible of 
the nuclear modifications of PDFs. Furthermore, they result in many free parameters.

A different approach to NPDFs was introduced in Refs.\cite{KP04, KP14}.
Nuclear PDFs are computed on the basis of an underlying microscopic model
incorporating several mechanisms of nuclear modifications   
including the smearing with the energy-momentum
distribution of bound nucleons (Fermi motion and binding),
the off-shell correction to bound nucleon PDFs,
the contributions from meson exchange currents and
the coherent propagation of the hadronic component
of the virtual intermediate boson in the nuclear environment.
This model explains to a high accuracy the observed
$x$, $Q^2$ and nuclear dependencies of the measured nuclear effects in DIS
on a wide range of targets from deuterium to lead~\cite{KP04,KP07,KP10}, as well as
the magnitude, the $x$ and mass dependence of all the available data from Drell-Yan
production off various nuclear targets~\cite{KP14}.

In this paper we perform a detailed study of the (pseudo)rapidity distributions of various
observables for $W^\pm$ and $Z^0$ productions in $p+{\rm Pb}$ collisions at the LHC
with the Kulagin and Petti (KP) nuclear PDFs~\cite{KP04, KP14}.
We compare our predictions with the 
recent CMS and ATLAS data at $\sqrt{s}=5.02$\,TeV and discuss the impact of
individual nuclear effects on the observed distributions.
To this end, the KP model allows an interpretation of the experimental results
in terms of the underlying nuclear physics mechanisms.
We also address the flavor dependence of the nuclear modifications of PDFs
in the context of both $W^+$ and $W^-$ distributions.
This topic is of particular interest since the CMS experiment reported
possible hints of such a flavor dependence from the $W$ charge asymmetry
measured in $p+{\rm Pb}$ collisions~\cite{Khachatryan:2015hha}.

This paper is organized as follows. In Sec.\ref{sec:wzdy} we outline the description of massive
vector boson production in the Drell-Yan process in proton-proton collisions. Section~\ref{sec:npdf}
summarizes the main features of the microscopic model used to calculate the KP nuclear PDFs.
In Sec.\ref{WZinpPb} we apply this model to study massive vector boson production in $p+{\rm Pb}$  
collisions at the LHC. Our results are presented in Sec.\ref{sec:res}, together with detailed
comparisons with the recent data from the CMS and ATLAS experiments at $\sqrt{s}=5.02$~TeV
at the LHC. In Sec.~\ref{sec:sum} we summarize.


\section{Vector Boson Production in the Drell-Yan Process}
\label{sec:wzdy}

The production of massive vector bosons ($W^\pm$ and $Z^0$, denoted as $V$)
through the DY mechanism in high-energy hadronic collisions is
a well understood process within the framework of
the perturbative QCD~\cite{Field:pQCD,Ellis:QCD}.
The QCD factorization theorem~\cite{Collins:1989gx} allows one to express the
corresponding production cross section as a convolution of the PDF
in the colliding hadrons with the partonic hard-scattering cross section,
which can be calculated in pQCD:
\begin{equation}
\frac{\ud\sigma^{DY}_{AB\rightarrow VX\rightarrow llX}}{\ud y}
=\sum_{a,b}\int \ud x_a \ud x_b q_{a/A}(x_a,Q^2)q_{b/B}(x_b,Q^2)
	\frac{\ud\hat{\sigma}_{ab\rightarrow VX\rightarrow llX}}{\ud y} .
\label{DY}
\end{equation}
where $q_{a/A}$ denotes the PDF of flavor $a$ in the hadron $A$,
the sum is taken over all possible parton flavors, and
$\ud\sigma_{AB}/\ud y$ and $\ud\hat{\sigma}_{ab}/\ud y$ are 
the hadronic and partonic differential cross sections as a function of the
vector boson rapidity $y$.
With the presence of a high-invariant-mass lepton pair $ll$ in the final state,
massive vector boson production provides a clean experimental signature to study the PDFs
of the hadrons involved in this process.
In this paper we discuss the vector boson production cross sections
focusing on $p+p$ and $p+{}^{208}\mathrm{Pb}$ collisions at the LHC.

The partonic cross section in \eq{DY}
can be calculated within pQCD order by order at the scale $Q^2$.
At the leading order in the $\alpha_S$ expansion (LO), this cross section is entirely
determined by the quark-antiquark annihilation process
(\eg, $u\bar{d}\rightarrow W^+\rightarrow l^+\nu$ or
$q\bar{q}\rightarrow Z^0\rightarrow l^+l^-$).
At the next-to-leading order (NLO), additional contributions to the partonic
cross section may arise from three different kinds of processes:
(i) one-loop virtual gluon corrections;
(ii) gluon emission corrections $q\bar{q}\rightarrow Vg$;
and (iii) the corrections from quark-gluon
scattering $qg\rightarrow Vq$ or $\bar qg\rightarrow V\bar q$~\cite{Ellis:QCD}.

The NLO and next-to-next-to-leading (NNLO) coefficients for the partonic cross sections of the
DY process and the hadronic $W$ and $Z$ boson production are well
known~\cite{Hamberg:1990np,Harlander:2002wh,Anastasiou:2003yy,Anastasiou:2003ds,Catani:2007vq,Catani:2009sm}.
It should be noted
that the NNLO corrections to the boson-rapidity distributions in relativistic heavy-ion collisions
are small at the LHC kinematics~\cite{Ru:2014yma,Ru:2015pfa}.

Our numerical analysis is carried out mostly to the NLO approximation in pQCD using the
DYNNLO program~\cite{Catani:2007vq,Catani:2009sm},
which is widely used to study the vector boson production at the LHC, as well as at the Tevatron.
As an essential input for our calculations, we use two different proton PDF sets:
ABMP15~\cite{Alekhin:2015cza} and
CT10~(2012 version)~\cite{Gao:2013xoa}.
The renormalization and the factorization scales are both set at the vector boson mass.

\begin{figure}[htb]
\centering
\includegraphics[width=\figx]{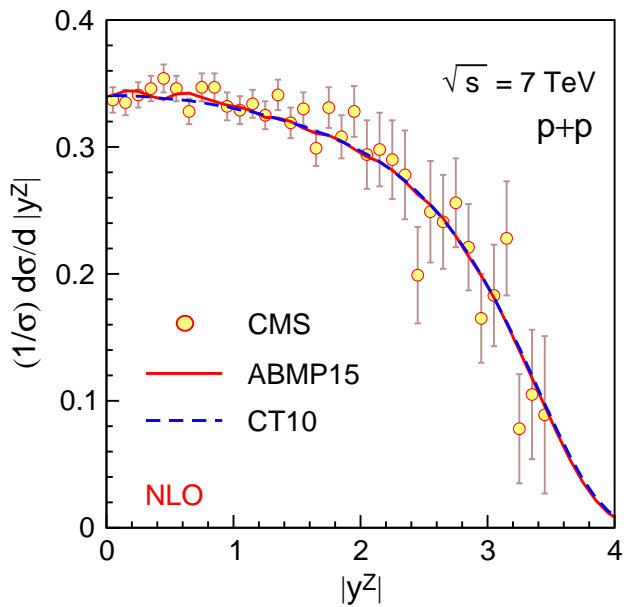}
\caption{Normalized differential cross section for $Z^0$ production in
$p+p$ collisions at $\sqrt{s}=7$~TeV as a function of the $Z^0$ rapidity.
The data points show the CMS measurement from Ref.~\cite{Chatrchyan:2011wt} with
statistical and systematic uncertainties added in quadrature. The 
invariant mass of the lepton pair is $60<m_{ll}<120$~GeV~\cite{Chatrchyan:2011wt}.
The curves are obtained from NLO calculations using two different PDF sets:
ABMP15~\cite{Alekhin:2015cza} (solid line) and CT10~\cite{Gao:2013xoa} (dashed line).
}
\label{Z_pp_CMS}
\end{figure}

Figure~\ref{Z_pp_CMS} shows the normalized differential cross section computed for 
$Z^0$ boson production in $p+p$ collisions at $\sqrt{s}=7$~TeV, as a function of
the $Z^0$ rapidity.
A good agreement between the NLO predictions and the CMS data~\cite{Chatrchyan:2011wt}
on $Z^0$ production is observed by using both the ABMP15 and the CT10 PDFs.
Similarly, the DYNNLO program provides a good description of the
$W$-boson production at the LHC~\cite{Ru:2014yma,Ru:2015pfa}.
%
%


\section{KP Nuclear Parton Distribution Functions}
\label{sec:npdf}

The calculation of the $W/Z$ production cross sections in $p+{\rm A}$ collision 
requires both proton and nuclear PDFs. 
In the present study we use the microscopic model of 
nuclear PDF of Refs.\cite{KP04,KP14} (KP model). 
In the following we will briefly summarize the main features of 
this model by using the DIS formalism in the nucleus rest frame 
for better clarity. However, we note that PDFs are universal Lorentz-invariant 
functions and therefore the results can be used to describe different processes 
like $W/Z$ production in any reference frame.   

The NPDF of Ref.\cite{KP14} include different contributions as follows: 
\begin{equation}
\label{npdf}
q_{a/A} = \left\langle q_{a/p}\right\rangle + \left\langle q_{a/n}\right\rangle
          + \delta q_a^\mathrm{MEC} + \delta q_a^\mathrm{coh} ,
\end{equation}
where $q_{a/A}$ is the PDF of flavor $a$ in a nucleus $A$
(for brevity we have suppressed the explicit dependencies on $x$ and $Q^2$).
The first two terms on the right side stand for the contribution from the
bound protons and neutrons,
and the brakets denote the averaging with the nuclear spectral function.
The terms $\delta q_a^\mathrm{MEC}$ and $\delta q_a^\mathrm{coh}$ are the corrections
arising from nuclear meson exchange currents (MEC) and the 
coherent interactions of the intermediate virtual boson with the nuclear target, respectively.

The first two terms in \eq{npdf} dominate in the valence region $x>0.1$
and in the nucleus rest frame can be written as a convolution with
the proton and neutron spectral function~\cite{Kulagin:1989mu,Kulagin:1994fz,KP04,KP14}.
In particular, for the proton contribution we have:
\begin{align}
\label{eq:IA}
\left\langle q_{a/p}\right\rangle =
\int\ud\varepsilon\ud^3{\bm p}\,
        \mathcal{P}_p(\varepsilon,{\bm p})\left(1+\frac{p_z}{M}\right)\frac{x'}{x}
        \, q_{a/p}(x',Q^2,p^2),
\end{align}
where the integration is taken over the energy $\varepsilon$
and the momentum $\bm p$ of the bound (off-shell) nucleon,
$\mathcal P_p$ is the spectral function describing the distribution over
energy and momentum of bound protons in the nucleus at rest, and $q_{a/p}$ is the PDF of the
bound proton with four-momentum $p=(M+\varepsilon,{\bm p})$ with $M$ being the proton mass.
The Bjorken variable of the nucleus is $x$
and the corresponding variable of the bound proton with four momentum $p$ is
$x'=Q^2/2p\cdot q=x/[1+(\ceps+p_z)/M]$
(the $z$ axis in \eq{eq:IA} is antiparallel to the direction of the momentum transfer $\bm q$).
A similar expression can be written for the bound neutron term in \eq{npdf}.
For brevity, we dropped $1/Q^2$ terms in \eq{eq:IA}
(for more detail see Ref.\cite{KP14}).

Note that \eq{eq:IA} was obtained starting from a Lorentz-covariant approach and
using a systematic expansion of matrix elements in series of the small parameters
$\bm p/M$ and $\ceps/M$, keeping terms of the order $\bm p^2/M^2$ and $\ceps/M$
\cite{Kulagin:1989mu,Kulagin:1994fz,KP04}.
The integrand in \eq{eq:IA} factorizes into two terms involving the contribution
from two different scales:
i)
the nuclear distribution $\mathcal P$ describing the processes at the nucleon level in the nuclear ground state,
and ii)
the PDF $q_{a/p}$ or $q_{a/n}$ describing the processes at the parton level in the nucleon.
The proton (neutron) spectral function in \eq{eq:IA} is normalized to the proton (neutron)
number in the nucleus.
This normalization condition also ensures the proper normalization of the
nuclear valence PDF by \eq{eq:IA}.
In applications we use a model spectral function, which includes both a mean field contribution
dominant at low energy and momentum,
and a high-momentum contribution related to short range nucleon-nucleon correlations~\cite{KP04}.

The off-shell nucleon PDF in \eq{eq:IA} explicitly depends on the nucleon invariant mass squared $p^2$.
Since the characteristic momenta of a bound nucleon are small compared to its mass,
the integration in \eq{eq:IA} mainly covers a region
in the vicinity of the mass shell and the nucleon virtuality $v=(p^2-M^2)/M^2$ 
can be considered a small parameter. We can then expand the PDF 
in series of $v$, keeping only the leading term~\cite{Kulagin:1994fz,KP04,KP14}:
\begin{align}
\label{SF:OS}
q_{a/p}(x,Q^2,p^2) &= q_{a/p}(x,Q^2)\left[ 1+\delta f(x,Q^2)\,v \right],
\\
\label{delf}
\delta f(x,Q^2) &= \partial\ln q_{a/p}(x,Q^2,p^2)/\partial\ln p^2,
\end{align}
where $q_{a/p}$ in the right side of \eq{SF:OS} is the PDF
of the on-mass-shell proton (or neutron) and the derivative is evaluated
on the mass shell $p^2=M^2$.

The off-shell (OS) function $\delta f$ can be regarded as a special nucleon structure function,
which describes the relative modification of nucleon PDF in the vicinity of the mass shell.
This function does not contribute to the cross sections of the
on-mass-shell nucleon, but it is relevant only for the bound nucleon and describes its
response to the interaction in a nucleus.
In general, the function $\delta f$ may depend on the PDF type and may be different
for protons and neutrons.
However, a detailed analysis of data on the ratios of DIS structure functions~\cite{KP04,KP10} 
and of DY cross sections \cite{KP14} for different nuclei 
supports the hypothesis of a universal OS function for all nucleon PDFs,
with no significant $Q^2$ dependence, \ie
$\delta f(x,Q^2)=\delta f(x)$.
The results of Ref.\cite{KP04} on $\delta f$ are also supported by
a recent combined analysis of DIS data off proton and deuteron targets,
Drell-Yan production in $pp$ and $pD$ interactions, and $W^\pm$ and $Z$ boson 
production in $pp$ and $p \bar p$ collisions \cite{AKP16}.
Therefore, we use a single universal off-shell function $\delta f(x)$ in computing all NPDFs.


The mesonic fields mediate the nucleon-nucleon interaction at distances exceeding
the typical nucleon size and also contribute to the quark-gluon content of the nucleus.
The nuclear correction $\delta q_a^\mathrm{MEC}$ in \eq{npdf}, originating from DIS off 
the virtual mesons exchanged between bound nucleons, 
can be written in terms of the convolution~\cite{KP04,KP14}:
\begin{equation}
\label{eq:MEC}
\delta q_a^\mathrm{MEC} = \sum_{m=\pi,\rho,\ldots} f_{m/A} \otimes q_{a/m}
\end{equation}
where the sum is taken over the possible meson states,  $f_{m/A}$ is the
light-cone distribution of the meson $m$ in the nucleus $A$, and $q_{a/m}$ is 
the parton distribution of flavor $a$ in the virtual meson $m$.
The meson light-cone distribution $f_{m/A}$ is calculated in Refs.\cite{KP04,KP14} by using
the nuclear light-cone momentum balance equation between bound nucleons and meson fields,
as well as the equation of motion for the meson fields.
We use the pion parton distribution functions from
Ref.\cite{Gluck:1999xe} to model the virtual meson PDF $q_{a/m}$ in \eq{eq:MEC}.
The MEC correction results in some enhancement of the nuclear sea-quark distribution
and its contribution is relevant in the region $x<p_F/M\sim 0.3$, where $p_F$ is the nuclear Fermi momentum.

The last term in \eq{npdf} is due to the propagation of the intermediate hadronic
states of a virtual boson in the nuclear environment.
We address this effect by replacing the sum over the set of all intermediate hadronic 
states by a single effective state and by describing 
its interaction with the nucleon with an effective scattering amplitude \cite{KP04}.
It is convenient to discuss coherent nuclear effects in terms of PDF combinations of definite
$C$-parity $q^\pm = q \pm \bar q$ and, for light quarks, of definite isospin
$q_0=u+d$ (isoscalar) and $q_1=u-d$ (isovector).
For example, for the $C$-even isoscalar PDF combination we have:
\begin{equation}
\label{eq:coh}
\delta q_0^\mathrm{coh}  =  q_{0/N} \, \Im \mathcal T^A(a_0^+)/\Im a_0^+,
\end{equation}
where $a_0^+$ is the $C$-even isoscalar forward effective scattering amplitude off the nucleon and
$\mathcal T^A$ is the sum of the nuclear multiple-scattering series for the effective nuclear 
amplitude in the corresponding channel.
A detailed discussion of other PDF combinations can be found in Ref.\cite{KP14}.

The term $\delta q_a^\mathrm{coh}$ is relevant at low $x$ and its
strength is governed by the effective amplitudes $a_I^C$
with different $C$-parity and isospin $I$.
In the region of small $x$ this correction is negative,
giving rise to the nuclear shadowing (NS) effect,
while in the transition region $x> 0.05$ the correction may be positive for some $I$ and $C$ channels,
because of a constructive interference between the amplitudes $a_I^C$ from different channels \cite{KP14}.

We note that different nuclear effects in different kinematical regions of
$x$ are related by the DIS sum rules and normalization constraints.
In Refs.\cite{KP04,KP14} these conditions are treated as dynamical constraints.
For example, as discussed above, the nuclear light-cone momentum sum rule at the hadronic level
(nucleons and mesons) links the nucleon and meson distribution functions.
The same sum rule at the partonic level constrains nuclear effects in the gluon distribution.
The normalizations of the isoscalar and the isovector valence quark distributions
(the baryon number and the Adler sum rules, respectively)
link the coherent and the off-shell corrections, since
the other contributions cancel out explicitly~\cite{KP14}.
In  Ref.\cite{KP04}, the off-shell effect provides an explicit mechanism to cancel a
negative nuclear shadowing contribution to the normalization of the nuclear valence quarks.
We also use the DIS sum rules to obtain the amplitudes $a_I^C$
in terms of the off-shell function $\delta f$ and the bound nucleon virtuality $v$ averaged
with the nuclear spectral function $\mathcal P$ in the corresponding isospin state $I$.

%
\begin{figure}[hbt]
\centering
\includegraphics[width=\figxx]{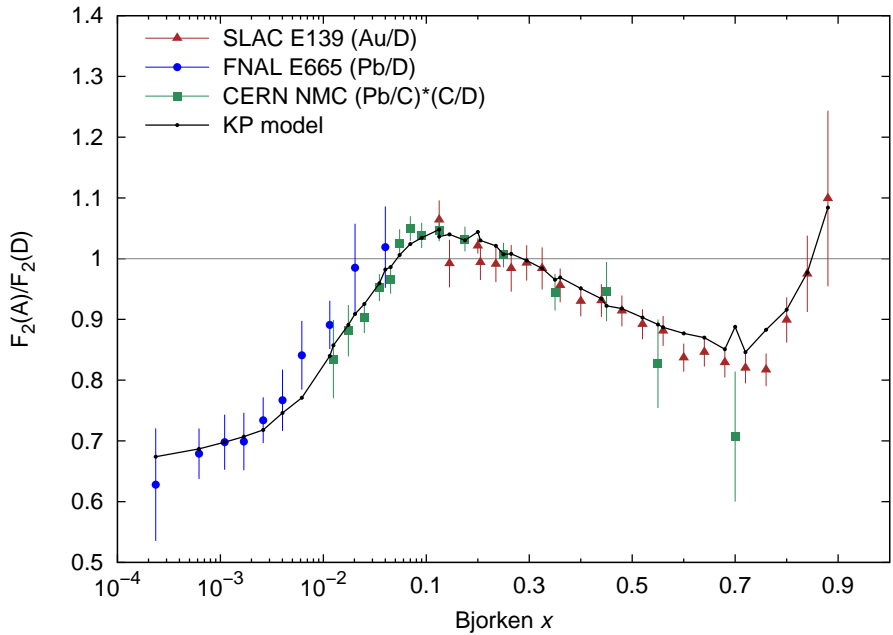}
\caption{%
Summary of data on $F_2(\mathrm{lead})/F_2(\mathrm{deuterium})$
and $F_2(\mathrm{gold})/F_2(\mathrm{deuterium})$
from the SLAC E139 \cite{e139}, FNAL E665 \cite{e665}, and CERN NMC \cite{nmc} experiments
(for NMC we show the product of the ratios lead/carbon and carbon/deuterium).
The dots connected by the solid line are the predictions of Ref.~\cite{KP04}
computed for the published values of $(x,Q^2)$ of each data point
(the wiggles are caused by different values of $Q^2$ for the CERN NMC and the SLAC E139 experiments).
We use a logarithmic scale for $x<0.1$ and a linear scale for $x>0.1$ for a 
better display of both the small $x$ and the large $x$ regions. 
}
\label{fig:disPbAu}
\end{figure}

A thorough analysis of data on the ratios of DIS structure
functions off different nuclei was carried out in Ref.\cite{KP04}
in the context of the described model.
The OS function $\delta f$, introduced in \eq{SF:OS},
was determined phenomenologically from this analysis with an approach similar 
to the one used for the other nucleon structure functions.
The model demonstrated an excellent performance and was able to describe
the observed $x$, $Q^2$ and $A$ dependencies of data to a high accuracy.
Figure~\ref{fig:disPbAu} summarizes the DIS data on ${}^{208}$Pb and ${}^{197}$Au
-- the nuclei relevant for the present study -- 
together with the corresponding predictions of Ref.\cite{KP04}.
The predictions of Ref.\cite{KP04} were further verified~\cite{KP10}
with the recent nuclear DIS data from
the HERMES experiment at HERA \cite{hermes} and the E03-103 experiment at JLab \cite{jlab}.
Furthermore, the same NPDF model describes well the
magnitude, the $x$ and mass dependence of the DY production cross section
off various nuclear targets~\cite{KP14} in the E772 \cite{e772} and E866 \cite{e866} experiments at Fermilab.

\begin{figure}[htb]
\includegraphics[width=\figx]{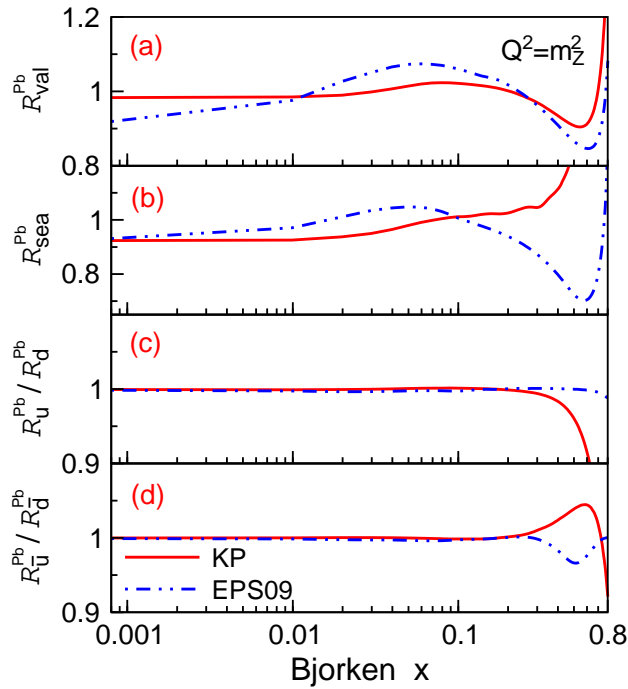}
\caption{%
Nuclear PDF ratios from \eq{eq:ra} computed following Ref.\cite{KP14} for $^{208}$Pb at $Q^2=m_Z^2$
as a function of the Bjorken $x$ (solid line). The different panels (top to bottom) show the nuclear corrections
for various PDF combinations:
(a) valence quark distributions;
(b) antiquark distributions;
(c) ratio $R^{Pb}_u/R^{Pb}_d$, and
(d) ratio $R^{Pb}_{\bar u}/R^{Pb}_{\bar d}$.
The results of Ref.\cite{Eskola:2009uj} are also shown for comparison.
}
\label{fig:nPDFs}
\end{figure}

Below we summarize briefly the main features of the KP nuclear PDFs.
For a nucleus of $Z$ protons and $N$ neutrons and $A=Z+N$ we define the ratio:
\begin{equation}
\label{eq:ra}
R_a^A(x,Q^2) = \frac{q_{a/A}(x,Q^2)}{Z q_{a/p}(x,Q^2)+N q_{a/n}(x,Q^2)},
\end{equation}
where $q_{a/A}$ is the nuclear PDF of flavor $a$,
and $q_{a/p}$ and $q_{a/n}$ are the corresponding PDFs for the free proton and neutron, respectively.
We assume the conventional isospin symmetry for the proton and neutron PDFs ($u_p=d_n$ and $d_p=u_n$).
Figure~\ref{fig:nPDFs} illustrates the ratios defined in \eq{eq:ra} for different
combinations of PDFs in the lead nucleus at $Q^2=m_Z^2$ (from top to bottom):
(a) nuclear correction $R^{\rm Pb}_{\rm val}$ for the valence quarks $u_v+d_v$;
(b) nuclear correction $R^{\rm Pb}_{\rm sea}$ for the full antiquark 
distribution $\bar u+\bar d+\bar s+\bar c+\bar b$;
(c) ratio $R^{\rm Pb}_u/R^{\rm Pb}_d$ related to the isospin-dependent nuclear effects on $u$ and $d$ quarks;
and (d) the ratio $R^{\rm Pb}_{\bar u}/R^{\rm Pb}_{\bar d}$ for the corresponding antiquarks.
For comparison we also show the corresponding nuclear correction ratios obtained from the
EPS09 phenomenological parametrization of NPDFs~\cite{Eskola:2009uj}.

Figure~\ref{fig:nPDFs} shows that the magnitude and the shape of the nuclear corrections is
different for the nuclear valence and sea-quark distributions.
In the region $x\ll 0.1$, the NPDFs are suppressed
by the nuclear \emph{shadowing} effect (negative $\delta q^{\rm coh}$ term).
However, the magnitude of this correction is not universal and differs for
the valence and sea quark distribution~\cite{KP04,KP14}.
We note that the result of the convolution of the nucleon PDF
with the nuclear spectral function in \eq{eq:IA} depends
upon the shape of the nucleon PDF considered.
In the small $x$ region this correction is positive for the valence
quarks, thus reducing the effect of the shadowing from the $\delta q^\mathrm{coh}$ term,
but is negative for the sea quarks.
However, the MEC correction for the sea quarks is positive, resulting in
a partial cancellation between different effects.
In the intermediate region $x\sim 0.1$
(usually referred as \emph{antishadowing} region)
we observe an interplay of different nuclear corrections.
For the the sea quark distributions we find an almost exact cancellation between
different nuclear corrections, while for the valence quarks we obtain a moderate enhancement,
which is caused by the interference between the $a_0^+$ and $a_0^-$ amplitudes in 
the multiple scattering correction \cite{KP04,KP14}.
At large $x>0.2$ nuclear PDFs are dominated by the incoherent scattering
from bound nucleons in \eq{eq:IA}.
For the valence quarks, the interplay between the
Fermi motion and nuclear binding correction (FMB) and
the off-shell correction results in a pronounced \emph{EMC-effect} at large $x$
\cite{Akulinichev:1985ij,Kulagin:1989mu,KP04}.
The relative size of this correction strongly depends on the particular $x$ dependence of the
input nucleon PDFs. For this reason the ratios $R_\mathrm{val}$ and $R_\mathrm{sea}$
are quite different at large $x$, as shown in Fig.\ref{fig:nPDFs}.


The last two panels in Fig.\ref{fig:nPDFs} illustrate the isospin (flavor) dependence of
the nuclear correction factors.
For the ratio $R_u/R_d$ the model of Ref.\cite{KP14} predicts
\begin{equation}
\label{R:ud}
\frac{R_u}{R_d} = 1 + 2\frac{Z{-}N}{A}\left(\frac{u-d}{u+d}\right)\frac{R_1-R_0}{R_0},
\end{equation}
where $u$ and $d$ represent the PDFs of the corresponding quarks at the given $(x,Q^2)$ kinematics,
and $R_0$ and $R_1$ are the nuclear corrections from \eq{eq:ra} for the isoscalar $q_0=u+d$
and the isovector $q_1=u-d$ PDF combinations, respectively.
A similar expression can be written for the double ratio of antiquark PDFs, $R_{\bar u}/R_{\bar d}$,
with $u$ and $d$ replaced by $\bar u$ and $\bar d$, respectively (for more details, see Ref.\cite{KP14}).
We note that the magnitude of $R_u/R_d{-}1$ appears to be significant only at large $x>0.3$,
where the last two factors become of the order of unity.

It is instructive to compare the nuclear effects on PDFs in Fig.\ref{fig:nPDFs}
with those on the structure function $F_2$ shown in Fig.\ref{fig:disPbAu}.
The most visible difference is a more pronounced shadowing correction at small $x$ for $F_2$ in DIS.
This effect can be explained with the strong $Q^2$ dependence of the effective cross section ($\Im a_0^+$)
describing the nuclear interaction in \eq{eq:coh}, since the results in Fig.\ref{fig:nPDFs} are obtained at
$Q^2=m_Z^2$, while the data in Fig.\ref{fig:disPbAu} have $Q^2\sim 1\gevsq$ or lower.
We also note that higher-twist terms (HT) as well as target mass corrections (TMC)~\cite{Georgi:1976ve}
play a significant role in the results shown in Fig.\ref{fig:disPbAu} (see Ref.\cite{KP04} for more details).


\section{$W^\pm/Z^0$ Production in $p+{\rm Pb}$ Collisions}
\label{WZinpPb}

The rapidity distributions of $W^\pm/Z^0$ bosons produced in nuclear collisions at the LHC offer
an excellent tool to study the cold nuclear medium effects on the (anti)quark
parton distributions~\cite{Vogt:2000hp,Paukkunen:2010qg,Guzey:2012jp,Ru:2014yma,Ru:2015pfa}.
The corresponding LO partonic processes are indeed initiated by quarks and antiquarks,
while gluons contribute only through higher order corrections to the subprocess cross sections.
The analysis of the nuclear modifications on the boson rapidity distributions in $p+{\rm Pb}$ collisions
is relatively simple compared to Pb+Pb collisions, since the nuclear partons are associated to a single nucleus
traveling in a definite direction.
Assuming the longitudinal axis along the proton beam direction,
the relations between the boson rapidity $y$ and the momentum fraction carried by
a parton from the proton or the lead nucleus can be written in the LO approximation as
\begin{equation}
\label{eq:y-x}
x_p=\frac{m_V}{\sqrt{s_{NN}}}e^{y},\quad\quad
x_{Pb}=\frac{m_V}{\sqrt{s_{NN}}}e^{-y}.
\end{equation}
From \eq{eq:y-x} one can estimate the typical momentum fraction carried by the nuclear partons
as $x_0=m_V/\sqrt{s_{NN}}\sim 0.017$ at the central rapidity $y \sim 0$.
If we neglect the contribution of nuclear gluons from higher order QCD corrections,
$x_0$ corresponds to the nuclear momentum fraction carried by (anti)quarks, with the
exception of the very forward region in which nuclear partons move rather slowly.
The kinematic region around $x_{Pb}\sim x_0$ corresponds to the transition between the
valence-dominated and the sea-dominated regions. As a result, both the valence and the sea quark distributions
play an important role in the study of $W^\pm/Z^0$ bosons produced in
nuclear collisions at the LHC~\cite{Ru:2014yma,Ru:2015pfa}.
Furthermore, bosons produced in the backward rapidity region ($y<0$, the direction of the lead beam) can provide
valuable information on the nuclear valence quark distributions,
while the nuclear sea quark distributions may play a dominant role in the nuclear modifications 
observed in the forward rapidity region ($y>0$, the direction of the proton beam).

Experimentally, it is easier to measure the pseudorapidity of the charged lepton originated
from the $W$ decay, $\eta^l$, rather than the $W$ boson rapidity, due to the additional smearing introduced by
the undetected neutrino present in the final state. The two variables are correlated and 
provide similar insights on the parton distributions~\cite{Paukkunen:2010qg}.

In addition to the differential cross sections of massive vector bosons,
observables defined as ratios of event rates are of particular interest,
like the forward-backward asymmetry $R_{FB}(y)$ and the $W$ charge asymmetry $\mathcal{A}(\eta^l)$:
\begin{align}
\label{FBasy}
R_{FB}(y) &= \frac{N(+y)}{N(-y)},
\\
\mathcal{A}(\eta^l) &= \frac{N^+(\eta^l)-N^-(\eta^l)}{N^+(\eta^l)+N^-(\eta^l)},
\label{Casy}
\end{align}
where $y$ is replaced with $\eta^l$ for $W$ boson production.
These ratios can enhance the sensitivity of various observables to the parton distributions and their nuclear
modifications, due to the partial cancellation of uncertainties in theoretical calculations
(\eg, scale dependence) and experimental measurements
(\eg, integrated luminosity)~\cite{Paukkunen:2010qg,Khachatryan:2015hha,Khachatryan:2015pzs}.
In particular, the $Z^0$ forward-backward asymmetry from \eq{FBasy}
is sensitive to the ratio of small-$x$ (sea-quark dominated) to large-$x$ (valence-quark dominated) NPDFs,
while the $W$ charge asymmetry in \eq{Casy} can shed light on the flavor dependence of the nuclear modifications
of PDFs (\eg, $R_u$ vs. $R_d$ and $R_{\bar u}$ vs. $R_{\bar d}$)~\cite{Khachatryan:2015hha}.

The discussion of $W^\pm$ and $Z$ production in $p+A$ collisions clearly requires one to address a 
number of cold nuclear matter effects affecting the PDFs of the colliding nucleus.
A standard approach to calculate the $W/Z$ production cross sections is to apply \eq{DY}
with the corresponding nuclear PDFs (NPDFs)
\cite{ConesadelValle:2009vp,Paukkunen:2010qg,Guzey:2012jp,Kang:2012am,%
Ru:2014yma,Ru:2015pfa,Albacete:2013ei,Albacete:2016veq,He:2011sg,Dai:2013xca,Neufeld:2010fj}.
A number of phenomenological NPDF parametrizations are available in the literature~\cite{Hirai:2007sx,
Eskola:2009uj,deFlorian:2011fp,Kovarik:2015cma,Khanpour:2016pph}.
In this work we use the KP NPDFs calculated on the basis of the microscopic model of Refs.~\cite{KP04,KP14},
which can provide a deeper understanding of the physics mechanisms responsible for the nuclear
modifications of massive vector boson productions in $p+A$ collisions (Sec.\ref{sec:npdf}).

In general, there is an interplay between the proton PDFs and the corresponding
nuclear corrections. In the KP model the convolution term by \eq{eq:IA} would result in
different nuclear correction factors for different input proton PDF.
For this reason the full calculation of nuclear PDFs described in Sec.\ref{sec:npdf}
has to be repeated when changing the set of proton PDFs.
The results shown in this paper are based on the five-flavor NNLO proton PDF set of Ref.\cite{Alekhin:2015cza}.
We note that the use of different proton PDFs requires some considerations even with other
NPDFs available in literature. Since NPDFs are typically determined from global QCD fits
to nuclear data, the use of a set of proton PDFs different from the one in the
corresponding QCD fits may result in violations of the valence quark normalizations
and momentum sum rule.
Another factor to consider are the corrections beyond the leading twist approximation,
such as the target mass correction (TMC) \cite{Georgi:1976ve} and the dynamical HT terms. 
These power corrections can significantly affect the NPDF analyses, which are 
dominated by the relatively low $Q^2$ DIS nuclear data.  
For instance it is known that TMC are sensitive to the shape of the proton PDFs used. 
\begin{figure}[htb]
\centering
\includegraphics[width=\figx]{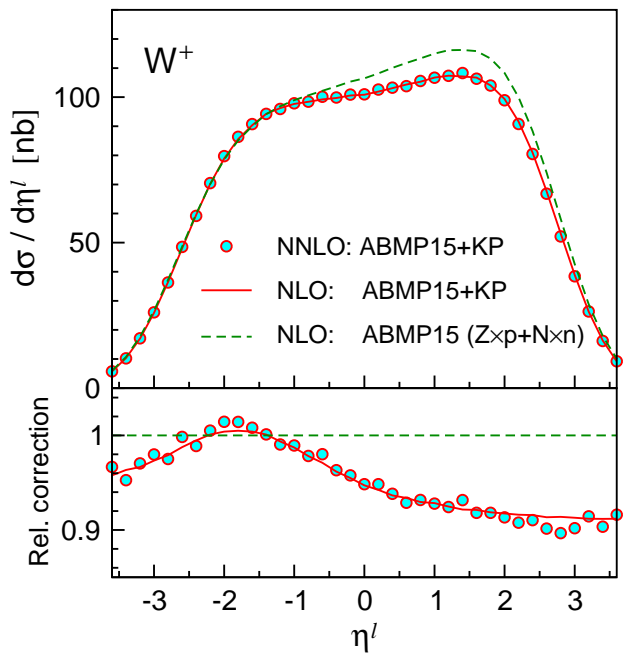}
\caption{Predictions for the differential cross section of $W^+$ production in $p+{\rm Pb}$ collisions at
$\sqrt{s_{NN}}=5.02$~TeV using different approximations: NNLO with ABMP15+KP (solid line),
NLO with ABMP15+KP (circles), and NLO with ABMP15 but with no nuclear
corrections (dashed line).
The bottom panel shows the corresponding ratios with respect to the results with no nuclear corrections.
}
\label{fig:nnlo}
\end{figure}

Figure~\ref{fig:nnlo} illustrates
the predictions for the $W^+$ differential cross section as a function of the
charged lepton pseudorapidity computed in different approximations.
From Fig.\ref{fig:nnlo} we conclude that the effect of NNLO correction on the partonic
cross sections is rather small, being even more marginal in the forward-backward asymmetry $R_{FB}(y)$ and
in the $W$ charge asymmetry $\mathcal{A}(\eta^l)$.
For this reason we use NLO partonic cross sections in the following analysis.

Figure\ref{fig:nnlo} clearly indicates the importance of nuclear corrections
in the process of $W/Z$ production in $p+{\rm Pb}$ collisions at the LHC energy. In Sec.\ref{sec:res} we perform
detailed comparisons of our predictions with the recent CMS data
on $W^\pm$ production~\cite{Khachatryan:2015hha} and $Z^0$ production~\cite{Khachatryan:2015pzs}
at $\sqrt{s}=5.02$\,TeV, as well as with the corresponding measurements
from the ATLAS experiment~\cite{ATLASW,Aad:2015gta}. For completeness,
our predictions are also compared with the results obtained using the EPS09 phenomenological
NPDF parametrization \cite{Eskola:2009uj}, supplemented by the CT10 proton PDFs \cite{Lai:2010vv}.
This choice is motivated by the fact that the CT10+EPS09 combination is widely used in the experimental
studies at the LHC, including the CMS and ATLAS measurements of $W^\pm/Z$ production in $p+{\rm Pb}$   
collisions of Refs.~\cite{Khachatryan:2015hha,Khachatryan:2015pzs,ATLASW,Aad:2015gta}.

\begin{figure*}[htb]
\includegraphics[width=\textwidth]{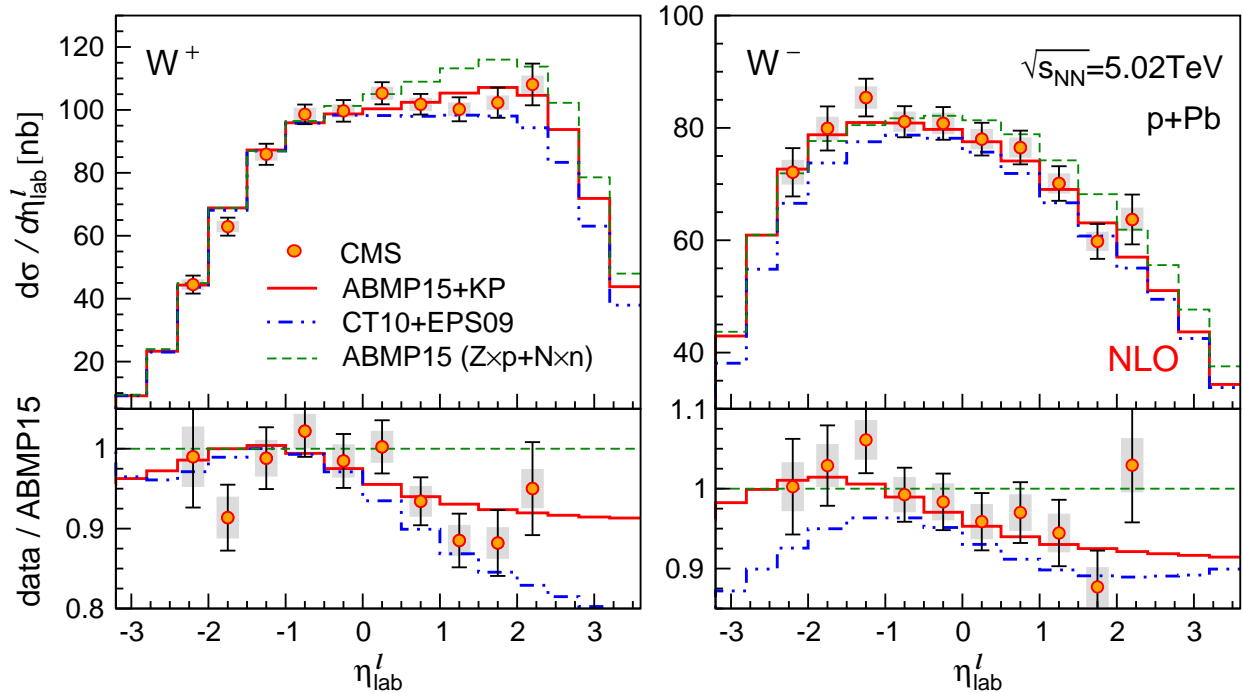}
\caption{%
Top panels: Differential cross sections for $W^+$ (left) and $W^-$ (right) production
in $p+{\rm Pb}$ collisions at $\sqrt{s_{NN}}=5.02$~TeV, as a function of the charged-lepton pseudorapidity.
The data points are the CMS measurement from Ref.~\cite{Khachatryan:2015hha} with statistical
uncertainties and total uncertainties (systematic and statistical uncertainties
added in quadrature) shown as gray boxes and solid bars, respectively.
The kinematic region covered corresponds to a charged-lepton transverse momentum 
$p_T^l>25$~GeV/c~\cite{Khachatryan:2015hha}. 
The curves represent the predictions computed using different models: ABMP15+KP (solid),
CT10+EPS09 (dash-dotted), and ABMP15 without nuclear modifications (dashed).
Bottom panels:
Ratios of the data (points with error bars) and the model predictions (curves) shown 
in the top panels with respect to the predictions with no nuclear corrections (ABMP15).
}
\label{fig:wcmseta}
\end{figure*}

\section{Results and Discussion}
\label{sec:res}

In Fig.\ref{fig:wcmseta} we compare our results on the differential cross sections
for $W^+$ and $W^-$ production in $p+{\rm Pb}$ collisions at $\sqrt{s_{NN}}=5.02$~TeV with the CMS
measurement from Ref.\cite{Khachatryan:2015hha}.
The cross sections are plotted as a function of the
charged-lepton pseudorapidity in the laboratory frame, $\eta_{lab}$.
For the CMS measurement \cite{Khachatryan:2015hha} the proton and the lead beam energies are
4\,TeV and 1.58\,TeV/nucleon, respectively.
Using these data, the relation between the pseudorapidity in the laboratory frame and that in the
center-of-mass frame can easily be calculated as $\eta_{lab}=\eta_{c.m.}+0.465$.

The precision of the recent CMS data provides some discriminating power among the theoretical predictions
obtained from ABMP15+KP, CT10+EPS09, and ABMP15 with no nuclear correction.
The results of the three calculations for the $W^+$ differential cross section
are consistent in the backward region, but display obvious differences in the forward region.
The CMS data clearly favor the presence of nuclear medium effects
as shown in Fig.\ref{fig:wcmseta}.
The predictions of different models for the $W^-$ differential cross section differ both
in the backward and in the forward regions
with the overall best description of the CMS data coming from the KP NPDFs.
The ratios in Fig.\ref{fig:wcmseta} indicate that the KP model predicts similar nuclear modifications
(suppression) for $W^+$ and $W^-$ in the forward region, but somewhat different corrections in the
backward region. This behavior can be explained with the flavor dependence of the nuclear modifications
in the KP model. As shown in Fig.\ref{fig:nPDFs}, the KP nuclear modifications on $u$ and $d$ quarks
are different in the large-$x$ (valence-quark dominated) region.
Differences are also present between nuclear $\bar{u}$ and $\bar{d}$.
Since the productions of $W^+$ and $W^-$ are dominated by different flavors
(\eg, $W^+$ by $u$ and $\bar{d}$, and $W^-$ by $d$ and $\bar{u}$), the corresponding rapidity distributions
are good observables to study the flavor dependence of nuclear modifications to PDFs.
We note that different nuclear corrections for valence and sea
quarks ($R_{val}$ vs. $R_{sea}$ in Fig.\ref{fig:nPDFs}) can also play a role in the $W^+$ and $W^-$
rapidity distributions, since the $W^+$ and $W^-$ production cross sections involve different 
fractions of valence (or sea) quarks.
For instance, at LO nuclear processes initiated by valence quarks contribute about
$65\%$ to the $W^+$ and $75\%$ to the $W^-$, at $\eta^l_{lab}\sim 2$.
The different behavior of nuclear modifications in the EPS09 and KP model shown in Fig.\ref{fig:nPDFs} 
plays a significant role in the difference observed on the $W^+$ and $W^-$ differential 
cross sections. 

\begin{figure*}[t]
\includegraphics[width=\textwidth]{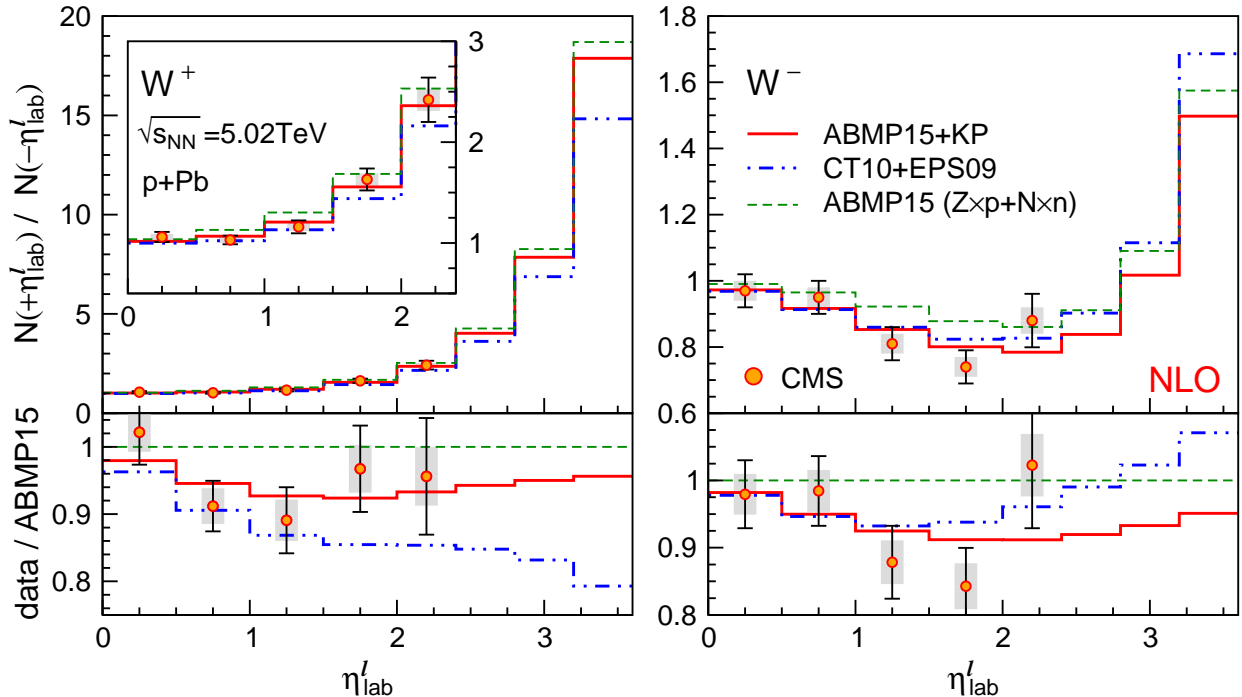}
\caption{Same notations as in Fig.\ref{fig:wcmseta} but for the lepton forward-backward asymmetry
$R_{FB}$ for $W^+$ (left) and $W^-$ (right) production in $p+{\rm Pb}$ collisions at $\sqrt{s_{NN}}=5.02$~TeV,
as a function of the charged lepton pseudorapidity.
}
\label{fig:wcmsfb}
\end{figure*}

Our results on the forward-backward asymmetry $R_{FB}$ for $W^\pm$ production, as a function of the
charged-lepton pseudorapidity in the laboratory frame, are illustrated in Fig.\ref{fig:wcmsfb}.
This observable offers a good sensitivity to nuclear modifications of PDFs
since, as discussed for the $W^\pm$ differential cross sections, the forward and backward regions 
are characterized by different nuclear corrections and parton content. 
The prediction with no nuclear modifications (ABMP15) does not reproduce well the $W^+$ nor the $W^-$ data.
Nuclear modifications are clearly needed to explain the general trend of the measured $R_{FB}$ distributions.

\begin{figure}[tbh]
\includegraphics[width=0.5\textwidth]{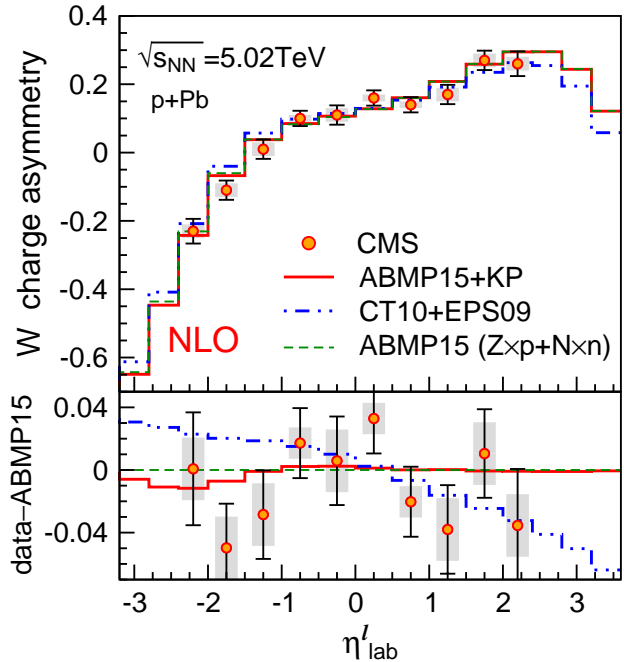}
\caption{Same notations as in Fig.\ref{fig:wcmseta} but for the charge asymmetry $\mathcal{A}$
of $W^\pm$ produced in $p+{\rm Pb}$ collisions at $\sqrt{s_{NN}}=5.02$~TeV,
as a function of the charged lepton pseudorapidity.
The lower panel shows the difference of data (points with error bars) and models (curves) 
indicated in the upper panel with respect to the predictions with no nuclear corrections (ABMP15).
}
\label{fig:wcmscas}
\end{figure}

In Fig.\ref{fig:wcmscas} we show our results for the $W$ charge asymmetry
as a function of the charged-lepton pseudorapidity in the laboratory frame.
The KP model predicts a small nuclear modification in the region
$-3< \eta^l_{lab}< -1.5$, due to the flavor dependence of the nuclear correction in the
valence-quark dominated region (Fig.~\ref{fig:nPDFs}) and partially to the different nuclear modifications
for valence and sea quarks, as discussed above.
The predictions with the KP NPDFs describe very well the CMS data over
the entire kinematic range. Similar results are obtained from the calculation
based upon the proton PDFs ABMP15.
Figure~\ref{fig:wcmscas} also indicates that the CT10+EPS09 model 
predicts a rather different shape for the charge asymmetry with respect to the ABMP15+KP model.
The expected values are systematically lower in the forward region and higher
in the backward region, resulting in a significant overestimation of the CMS data in
the region $-2\lesssim \eta^l_{lab}\lesssim -1$~\cite{Khachatryan:2015hha}.
As shown in Fig.\ref{fig:nPDFs}, no significant flavor dependence is present in the EPS09 
corrections, due to the initial assumption of isospin symmetry $R_u = R_d$.
We note that a large part of the differences between the CT10+EPS09 and ABMP15+KP 
curves is related to the underlying proton PDFs used since the effect of nuclear corrections
is reduced in the $W$-boson charged asymmetry. This partial cancellation is visible 
from a comparison of the curves obtained with ABMP15+KP and ABMP15 only
in Fig.~\ref{fig:wcmscas}. For the effect of different proton PDFs in EPS09
see also Ref.~\cite{Armesto:2015lrg}.
The backward region is dominated by the valence quarks in the lead nucleus, while the forward region
is related to the large-$x$ partons in the forward going proton.
Therefore, the $u/d$ proton PDF ratio at large $x$ is particularly relevant for the $W$ charge asymmetry,
as well as the $\bar{u}/\bar{d}$ ratio at small $x$.
\begin{figure*}[thb]
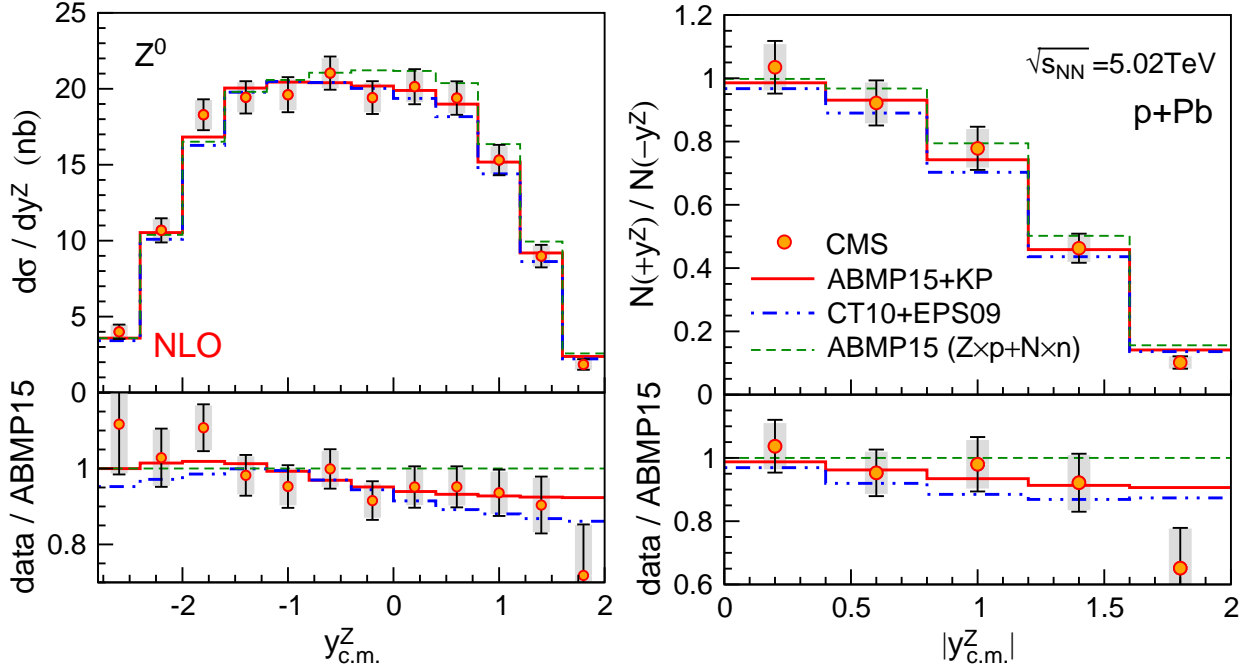

\includegraphics[scale=0.78]{z0cms5ppbnloyu.eps}
\includegraphics[scale=0.78]{z0cms5ppbfbasyu.eps}
\caption{Same notations as in Fig.\ref{fig:wcmseta} but for
the differential cross section (left) and the related forward-backward asymmetry $R_{FB}$  (right)
for $Z^0$ production in $p+{\rm Pb}$ collisions at $\sqrt{s_{NN}}=5.02$~TeV,
as a function of the $Z^0$ rapidity.
The data points are the CMS measurement \cite{Khachatryan:2015pzs}.
The kinematic region covered corresponds to a lepton pair invariant mass 
$60<m_{ll}<120$~GeV and a lepton transverse momentum $p_T^l>20$~GeV/c, 
leading to $|\eta_{lab}^l|<2.4$~\cite{Khachatryan:2015pzs}.
}
\label{fig:zcms}
\end{figure*}

Figure~\ref{fig:zcms} summarizes our results on the differential cross section and the
corresponding forward-backward asymmetry for $Z^0$ production in $p+{\rm Pb}$ collisions at
$\sqrt{s_{NN}}=5.02$~TeV, as a function of the $Z^0$ rapidity in the
center-of-mass frame of the nucleon-nucleon collision.
In particular, the $Z^0$ forward-backward asymmetry $R_{FB}$ offers a clean probe
for the study of cold medium nuclear effects~\cite{Paukkunen:2010qg}.
The KP nuclear modifications suppress the rate of $Z^0$ production in the forward rapidity region and
slightly enhances it in the backward rapidity region ($-2.5< y^Z< -1.2$).
The resulting forward-backward asymmetry is therefore suppressed, similarly to the case
of $W^\pm$ production.
The KP model predictions are in excellent agreement with the CMS data for both the differential cross section
and the forward-backward asymmetry.
Figure~\ref{fig:zcms} shows that the results based on the ABMP15+KP model
and the CT10+EPS09 parametrization are somewhat different.
The difference is mainly related to the corresponding nuclear modification factors of PDFs,
since our results for $Z^0$ production in $p+p$ collisions (see Fig.\ref{Z_pp_CMS}) indicate
that the ABMP15 and CT10 predictions are consistent.
For the parton kinematics associated to the backward rapidity region, $0.02< x_{Pb} < 0.1$,
the EPS09 nuclear modifications at $Q^2=m_Z^2$ lead to a stronger enhancement than the KP model for both the
valence and sea quark distributions (see Fig.\ref{fig:nPDFs}).
Similarly, in the region $0.001< x_{Pb} < 0.01$, corresponding to the forward direction,
the EPS09 introduces a stronger suppression of the valence quarks (see also Fig.\ref{fig:nPDFs}).
As a result, the EPS09 predicts somewhat lower values for the forward-backward asymmetry
with respect to the KP model.

\begin{figure*}[htb]
\includegraphics[width=\textwidth]{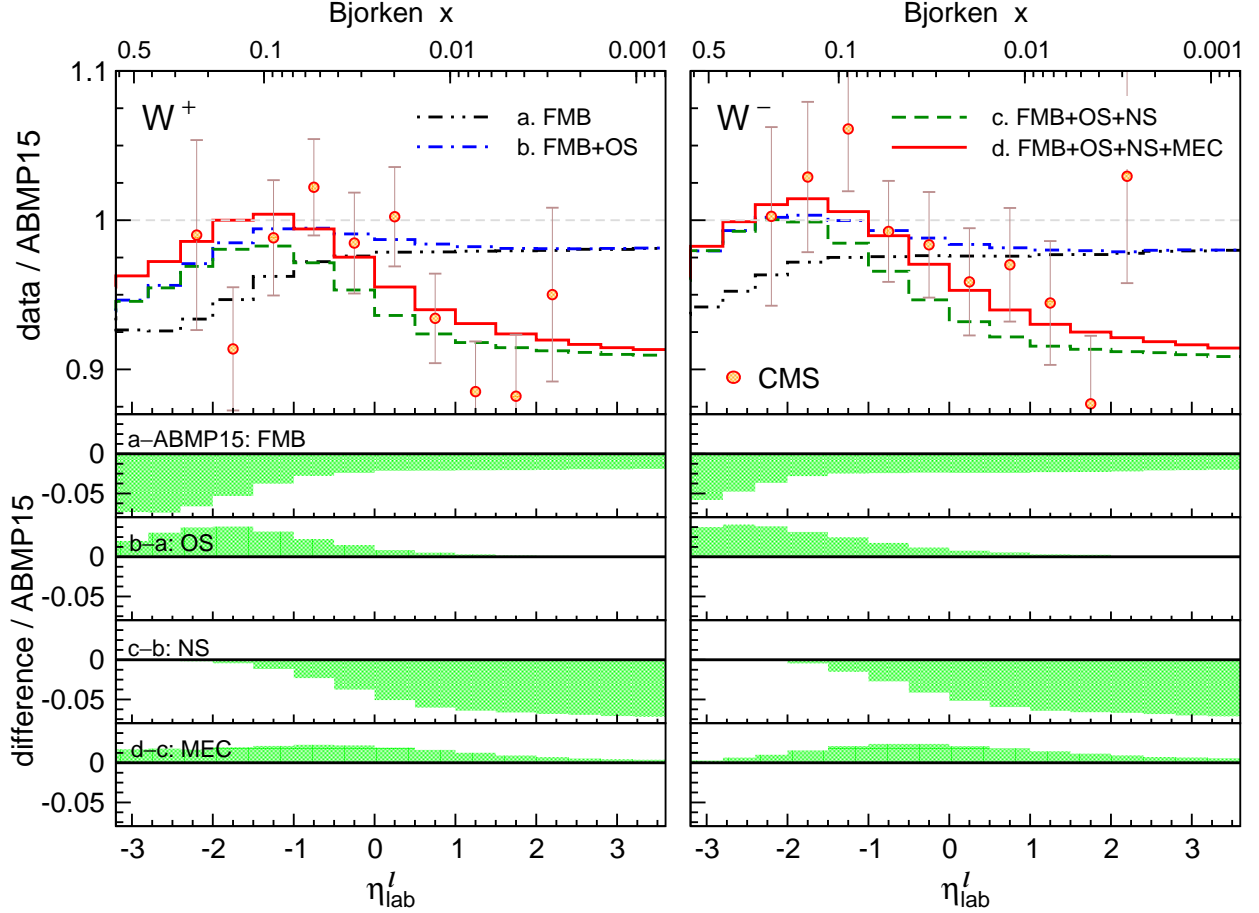}
\caption{Top panels: Nuclear corrections calculated from the ratios defined in \eq{eq:RpPb} for the
differential cross sections of $W^+$ (left) and $W^-$ (right) production in $p+{\rm Pb}$ collisions at
$\sqrt{s_{NN}}=5.02$~TeV. The curves illustrate the impact of adding different cold nuclear matter effects
in the KP model (see text for details):
FMB (dash-dot-dotted),
FMB+OS (dash-dotted),
FMB+OS+NS (dashed),
and the full calculation FMB+OS+NS+MEC (solid).
The data points from the CMS measurement \cite{Khachatryan:2015hha} are also shown for
comparison (the error bars correspond to the sum in quadrature of statistical and systematic uncertainties).
A double horizontal scale is used for completeness: the bottom one shows the charged lepton pseudorapidity,
while the top one provides an estimate of the equivalent Bjorken $x_{Pb}$ for the partons in the
lead nucleus.
Bottom panels: Relative contribution of each individual nuclear effect on the
nuclear corrections for the $W^+$ (left) and $W^-$ (right) differential cross sections.
Each contribution is obtained by subtracting the corresponding curves in the top panels,
with and without the effect considered.
}
\label{fig:wcmskp}
\end{figure*}

One advantage of the KP nuclear PDFs is that they are based upon a detailed microscopic model 
(see Sec.~\ref{sec:npdf}) allowing one to disentangle the contributions from different 
mechanisms responsible for the nuclear modification of PDFs. 
In order to discuss the sensitivity of CMS data to individual nuclear effects we define
the KP nuclear modification ratio for the $W/Z$ differential cross sections as:
\begin{equation}
\mathcal{R}^{\mathrm{KP}}_{pPb}(\eta^l)=\frac{(\ud\sigma/\ud\eta^l)_{\mathrm{KP}}}{(\ud\sigma/\ud\eta^l)_{\mathrm{ABMP15}}},
\label{eq:RpPb}
\end{equation}
where $\eta^l$ should be replaced by $y^Z$ for $Z^0$ production.
We evaluate this ratio using different combinations of nuclear effects in the KP model, as
summarized in Sec.\ref{sec:npdf}:
(a) Fermi motion and binding correction (FMB) only;
(b) FMB+ off-shell correction (OS);
(c) FMB+OS+ coherent corrections related to nuclear shadowing (NS);
and (d) the complete model FMB+OS+NS+ meson exchange currents (MEC).
Results are shown in Fig.\ref{fig:wcmskp} for the $W^+$ and $W^-$ differential cross sections
in $p+{\rm Pb}$ collisions at $\sqrt{s_{NN}}=5.02$~TeV, 
together with the corresponding CMS data~\cite{Khachatryan:2015hha}.
For a better understanding of the various nuclear effects at the parton level in the upper scale
we also show the values of the Bjorken variable $x_{Pb}$ obtained from \eq{eq:y-x}.
The relative impact of each individual nuclear effect on the cross sections can be evaluated from the
difference of the ratios defined in \eq{eq:RpPb} with and without the effect considered, as
shown in Fig.\ref{fig:wcmskp}.

From Fig.\ref{fig:wcmskp} we can observe that the kinematical coverage of $W^\pm$ production 
in $p+{\rm Pb}$ collisions in the CMS experiment is sensitive to all four physics mechanisms
responsible for the nuclear modification of PDFs. A comparison with Fig.\ref{fig:disPbAu}
shows that in the probed region of the Bjorken $x$ we
expect significant variations in the nuclear corrections.
While the size of the combined effect of FMB+OS at large $x$ is comparable in DIS and
$W^\pm$ production, the shadowing correction in Fig.~\ref{fig:wcmskp} appears to be substantially
reduced with respect to the nuclear DIS.
This difference can be attributed to the fact that the $Q^2$ scale differs by 4 orders of magnitude
($Q^2\sim 1\gevsq$ for fixed-target DIS in Fig.~\ref{fig:disPbAu} and $Q^2\sim 10^4\gevsq$ for Fig.\ref{fig:nPDFs})
and the corresponding effective cross sections driving the shadowing corrections at small $x$ 
(see Sec.\ref{sec:npdf}) are significantly different \cite{KP14}.
As discussed in Sec.\ref{sec:npdf}, significant high twist contributions are also present in the  
low $Q^2$ DIS data shown in Fig.~\ref{fig:disPbAu}.  

\begin{figure}[htb]
\includegraphics[width=0.5\textwidth]{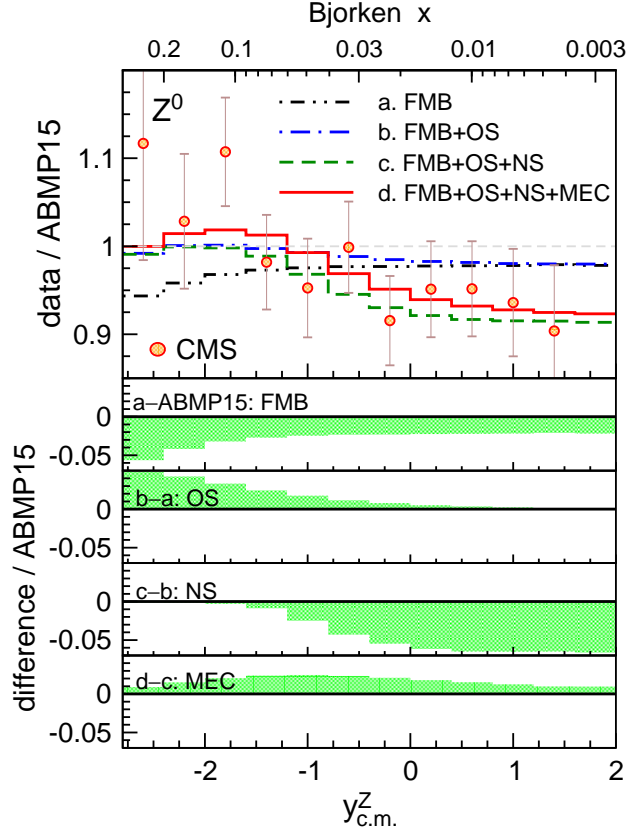}
\caption{Same notations as in Fig.\ref{fig:wcmskp} but for the $Z^0$ differential cross section
as a function of the $Z^0$ rapidity.
The data points indicate the CMS measurement \cite{Khachatryan:2015pzs}.
}
\label{fig:zcmsKP}
\end{figure}

The FMB and OS corrections dominate the backward region $\eta^l_{lab}< -1.5$,
mainly due to their effect in the valence-quark region $x_{Pb}> 0.1$.
In this rapidity region the FMB correction is negative (suppression),
while the corresponding OS correction is positive (enhancement), for both $W^+$ and $W^-$ production.
In the forward region Fig.\ref{fig:wcmskp} shows a suppression as a result of the nuclear
shadowing on small-$x$ partons.
The enhancement observed in the intermediate and backward regions can be related to the
nuclear meson correction, affecting the nuclear sea quark distributions (mainly $u$ and $d$)
for $x< 0.2$.
It is worth noting that the shadowing corrections on $W^+$ and $W^-$ production appear to
be similar. Instead, differences between $W^+$ and $W^-$ are observed in the nuclear corrections
originated by the other physics mechanisms as a consequence of their flavor dependence
($R_u\neq R_d$, and $R_{val}\neq R_{sea}$ in Fig.\ref{fig:nPDFs}).

Figure~\ref{fig:wcmskp} indicates that the CMS data are rather sensitive to the
off-shell correction. As discussed in Sec.\ref{sec:npdf}, the off-shell effect plays an
important role in the KP model (together with the FMB),
through the off-shell structure function $\delta f$ in \eq{SF:OS}.
The predictions of the KP model for this study assume a single universal off-shell
function for all PDFs. However, in general, this function may be flavor dependent and different
for bound protons and neutrons. A comparison of $W^+$ and $W^-$ production in $p+{\rm Pb}$  
collisions can potentially shed some light on these issues. The current CMS data
are consistent with the assumption of a universal function, but future high precision data
would be very valuable to further clarify this point.

\begin{figure*}[t]
\includegraphics[scale=0.65]{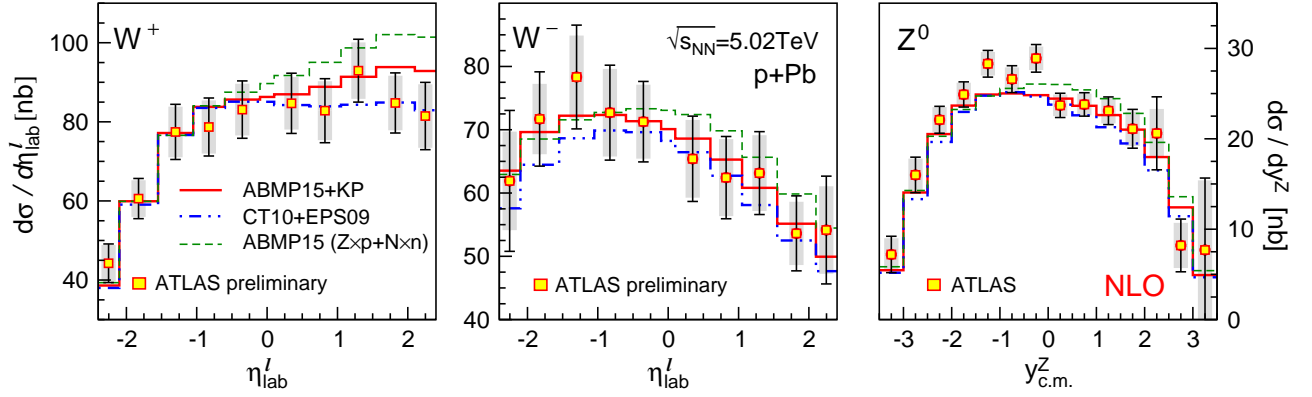}
\caption{Differential cross sections for $W^+$ (left), $W^-$ (middle), and $Z^0$ (right) production
in $p+{\rm Pb}$ collisions at $\sqrt{s_{NN}}=5.02$~TeV, as a function of the (pseudo)rapidity.
The data points indicate the ATLAS measurement of $W^\pm$ production (preliminary)~\cite{ATLASW}
and the $Z^0$ production~\cite{Aad:2015gta}
with statistical and total uncertainties (systematic and statistical uncertainties
added in quadrature) shown as gray boxes and solid bars, respectively.
The kinematic region covered corresponds to a muon preudorapity $0.1<|\eta_{lab}^l|<2.4$, 
a muon transverse momentum $p_T^l>25$~GeV/c, a neutrino transverse momentum $p_T^{\nu}>25$~GeV/c, 
a transverse mass $m_T>40$~GeV~\cite{ATLASW}, and a $Z$ boson invariant mass 
$66<m_{ll}<116$~GeV~\cite{Aad:2015gta}.
The curves are the predictions based on different models: ABMP15+KP (solid),
CT10+EPS09 (dash-dotted), and ABMP15 with no nuclear modifications (dashed).
}
\label{fig:atlaswz}
\end{figure*}

In Fig.\ref{fig:zcmsKP} we show the contributions from different nuclear effects to the
$Z^0$ differential cross section in $p+{\rm Pb}$ collisions at $\sqrt{s_{NN}}=5.02$~TeV,
together with the corresponding CMS data~\cite{Khachatryan:2015pzs}.
Similar considerations can be made as for the $W^\pm$ cross sections in
Fig.\ref{fig:wcmskp}.

In the previous discussion we mainly focused on the various observables from the recent measurements
by the CMS experiment. However, the ATLAS experiment also measured the
$W/Z$~\cite{ATLASW,Aad:2015gta} rapidity distributions in $p+{\rm Pb}$ collisions at $\sqrt{s_{NN}}=5.02$~TeV,
although the $W^\pm$ data are still preliminary~\cite{ATLASW}.
For completeness, we calculate the predictions for the differential cross sections of
$W/Z$ production in ATLAS and compare them with the available data in Fig.\ref{fig:atlaswz}.
The differences among the predictions of ABMP15+KP, CT10+EPS09, and ABMP15 with no nuclear effects
for $W^\pm$ and $Z^0$ production in ATLAS are very similar to those discussed in the CMS context.
Overall, the ABMP15+KP model predictions describe well the ATLAS data.
For the $Z^0$ production, we observe a small excess in the data points at
$-2< y < 0$, which is not present in the CMS data shown in Fig.\ref{fig:zcms}.

\begin{figure*}[htb]
\centering
\includegraphics[scale=0.6]{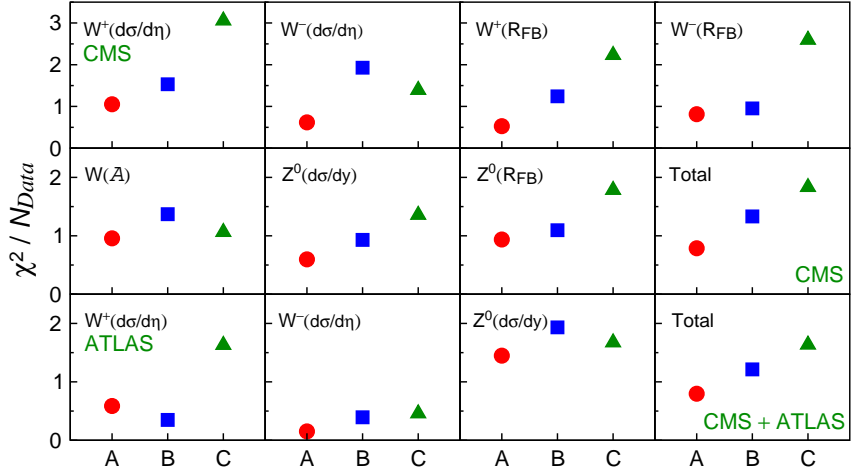}
\caption{Normalized $\chi^2$ (per data point) for the various observables
shown in the previous plots, calculated between each data set and three different
model predictions: ABMP15+KP (A), CT10+EPS09 (B), and ABMP15 with no nuclear corrections (C).
See Table~\ref{tab:chisqlhc} for more details.
}
\label{fig:chisqlhc}
\end{figure*}
\begin{table}[hbt]
\caption{Normalized $\chi^2$ (per data point) for the various observables (rows)
shown in the plots listed in the first column, calculated between each data set and three different
model predictions: ABMP15+KP, CT10+EPS09, and ABMP15 with no nuclear corrections (last column).
}
\begin{center}
\begin{tabular}{c|c|ccc}
 \hline
  {\rm \textsf{Observable}} & $N_{\rm Data}$ & ~~~\textsf{ABMP15}~~~ & ~~\textsf{CT10} & \textsf{ABMP15}\\
  &  &~~~+ \textsf{KP}~~~ & ~~~+ \textsf{EPS09}~~~  & $~~~{(\textsf{Zp} +\textsf{Nn})}~~~$  \\
\hline\hline
   &   & \multicolumn{3}{c}{CMS experiment:} \\
  $\textsf{d}\sigma^+\textsf{/d}\eta^l$ & 10 & \textsf{1.052} & \textsf{1.532} & \textsf{3.057}\\
  $\textsf{d}\sigma^-\textsf{/d}\eta^l$ & 10 & \textsf{0.617} & \textsf{1.928} & \textsf{1.393}\\
  \scriptsize$\textsf{N}^+\textsf{(}+\eta^l\textsf{)}/\textsf{N}^+\textsf{(}-\eta^l\textsf{)}$ & 5 & \textsf{0.528} & \textsf{1.243} & \textsf{2.231}\\
  \scriptsize$\textsf{N}^-\textsf{(}+\eta^l\textsf{)}/\textsf{N}^-\textsf{(}-\eta^l\textsf{)}$ & 5 & \textsf{0.813} & \textsf{0.953} & \textsf{2.595}\\
   \scriptsize$\textsf{(N}^+-\textsf{N}^-\textsf{)/(N}^++\textsf{N}^-\textsf{)}$ & 10 & \textsf{0.956} & \textsf{1.370} & \textsf{1.064}\\
   $\textsf{d}\sigma\textsf{/dy}^Z$ & 12 & \textsf{0.596} &\textsf{0.930} & \textsf{1.357}\\
   \scriptsize$\textsf{N(}+\textsf{y}^Z)\textsf{/N(}-\textsf{y}^Z)$ & 5 & \textsf{0.936} & \textsf{1.096} & \textsf{1.785}\\ \hline
   \textsf{CMS combined} & 57 & \textsf{0.786} &\textsf{1.332} & \textsf{1.833}\\
   \hline
   &  & \multicolumn{3}{c}{ATLAS experiment:} \\
  $\textsf{d}\sigma^+\textsf{/d}\eta^l$ & 10 & \textsf{0.586} & \textsf{0.348} & \textsf{1.631} \\
  $\textsf{d}\sigma^-\textsf{/d}\eta^l$ & 10 & \textsf{0.151} & \textsf{0.394}  &\textsf{0.459} \\
  $\textsf{d}\sigma\textsf{/dy}^Z$ & 14 & \textsf{1.449} & \textsf{1.933}  &\textsf{1.674} \\
\hline
  \textsf{CMS+ATLAS combined} & 91 & \textsf{0.796} & \textsf{1.213} & \textsf{1.635}\\
 \hline
\end{tabular}
\end{center}
\label{tab:chisqlhc}
\end{table}

In order to make quantitative comparisons between the various predictions and the available data,
we evaluate the normalized $\chi^2$ for each experimental observable as:
\begin{equation}
\chi^2/N_{Data}=\frac{1}{N_{Data}}{\sum^{N_{Data}}_{i=1}{\left[\frac{\left(O_{\rm th}-O_{\rm exp}\right)^2}{\varepsilon^2_{stat}+\varepsilon^2_{syst}}\right]_i}},
\end{equation}
where $O_{\rm th}$ and $O_{\rm exp}$ are the theoretical prediction and the experimental measurement
for the $i$-th data point, respectively,
and $\varepsilon_{stat}$ and $\varepsilon_{syst}$ are the corresponding statistical and
systematic uncertainties.
The results obtained for the different models are summarized in
Fig.\ref{fig:chisqlhc} and listed in Table~\ref{tab:chisqlhc}.
A comparison between the normalized $\chi^2$ values obtained with the KP NPDFs and the
ones obtained without nuclear corrections (ABMP15 only) clearly shows the importance of nuclear
modifications of PDFs for both CMS and ATLAS data. This observation can be
interpreted as evidence for the presence of nuclear effects in $W/Z$ production in $p+{\rm Pb}$ collisions.
The predictions with KP nuclear PDFs provide the best description of both CMS and ATLAS data,
with an overall value of $\chi^2/N_{Data}=0.796$ for the combined CMS+ATLAS data set
with $N_{Data}=91$.
This result demonstrates that the KP nuclear PDFs can be a powerful tool in the study
of hard scattering processes in heavy-ion nuclear collisions.
It will be interesting to extend our analysis with the KP NPDFs
to other physics observables in hard scattering processes such as direct photon
production~\cite{Dai:2013xca}, hadron production at large transverse momentum~\cite{Dai:2015dxa},
inclusive jet~\cite{Vitev:2009rd} and dijet productions~\cite{He:2011sg},
as well as gauge bosons tagged jet productions~\cite{Neufeld:2010fj,Dai:2012am} in both 
$p+{\rm Pb}$ and ${\rm Pb}+{\rm Pb}$ collisions.
Such studies will allow one to understand how different nuclear matter effects are constrained by
existing experimental measurements and to achieve a more robust separation between the
initial-state cold nuclear matter effects and the final-state hot quark-gluon-plasma medium effects
in relativistic heavy-ion collisions~\cite{Albacete:2013ei,Albacete:2016veq}.


\section{Summary}
\label{sec:sum}

We performed a detailed study of the (pseudo)rapidity distributions of various
observables for $W/Z$ productions in $p+{\rm Pb}$ collisions with
$\sqrt{s}=5.02$\,TeV at the LHC, using the KP nuclear PDFs together with the DYNNLO program.
In this approach the nuclear modifications are computed
from an underlying microscopic model including several nuclear physics mechanisms
including nuclear Fermi motion and binding, off-shell correction to bound nucleon PDFs, meson
exchange currents in nuclei, and coherent effects responsible for the nuclear shadowing.

We performed a detailed comparison between the model predictions and
the recent precision data on $W^\pm$ and $Z^0$ productions in $p+{\rm Pb}$ collisions
from the CMS and ATLAS experiments at the LHC.
The data clearly favor the presence of nuclear modifications on the $W/Z$ production
cross sections with respect to the case of $p+p$ collisions.
We found an excellent agreement between the predictions based on the KP NPDFs
and all the measured observables in the
entire kinematic range accessible by the experiments.
Our analysis of CMS and ATLAS data showed that the KP model 
can provide interesting insights on the underlying physics mechanisms
responsible for the nuclear modifications of PDFs.
 
We found that the kinematics coverage of $W/Z$ production in $p+{\rm Pb}$ collisions in the
CMS and ATLAS experiments is sensitive to all underlying nuclear effects responsible
for the nuclear modifications of PDFs in the KP model.
For this reason, the full nuclear correction on $W/Z$ production in $p+{\rm Pb}$  
collisions is the result of an interplay of different physics mechanisms.
We also discussed the flavor dependence of the nuclear correction with a detailed
analysis of both $W^+$ and $W^-$ distributions. In particular, we found that
the KP model can correctly describe the $W$ charge asymmetry
reported by the CMS experiment in $p+{\rm Pb}$ collisions.
 
Finally, we note that the precision currently achieved by the LHC experiments --
most notably with the latest CMS measurements of $W^\pm/Z$ production --
starts to be sensitive to the predicted nuclear corrections.
A further improvement of the accuracy of future data sets would be extremely valuable in this context
since it could allow to disentangle the effect of different underlying mechanisms responsible for
the nuclear modifications of PDFs and to study their flavor dependence.

\section*{Acknowledgments}

We thank S. Alekhin for providing the ABMP15 parametrization of the proton PDFs.
We thank H. Paukkunen for comments on the manuscript. 
P.R. and B.W.Z. were supported by the Ministry of Science and Technology of 
China under Projects No. 2014CB845404, No. 2014DFG02050, and
by the Natural Science Foundation of China with Projects No. 11322546, No. 11435004, and No. 11521064.
R.P. was supported by the Grant No. DE-SC0010073 from the Department of Energy, U.S.A.
S.K. was supported by the Russian Science Foundation Grant No. 14-22-00161.


\end{document}